\documentstyle[preprint,graphics,aps]{revtex}
\begin{document}
\draft
\title{A Study of the Decay $D^0\!\rightarrow\!K^+\pi^-$}

\author{
J.M.~Link$^{1}$,
M.~Reyes$^{1}$,
P.M.~Yager$^{1}$,
J.C.~Anjos$^{2}$,
I.~Bediaga$^{2}$,
C.~G\"obel$^{2}$,
J.~Magnin$^{2}$,
A.~Massafferi$^{2}$,
J.M.~de~Miranda$^{2}$,
I.M.~Pepe$^{2}$
% ...
}

\author{(The FOCUS Collaboration)}

\address{$^1$University of California, Davis, CA 95616}
\address{$^2$Centro Brasileiro de Pesquisas F\'\i sicas, Rio de Janeiro,
RJ, Brazil}
\address{$^3$CINVESTAV, 07000 M\'exico City, DF, Mexico}
\address{$^4$University of Colorado, Boulder, CO 80309}
\address{$^5$Fermi National Accelerator Laboratory, Batavia, IL 60510}
\address{$^6$Laboratori  Nazionali di Frascati dell'INFN, Frascati, Italy, 
I-00044}
\address{$^7$University of Illinois, Urbana-Champaign, IL 61801}
\address{$^8$Indiana University, Bloomington, IN 47405}
\address{$^9$Korea University, Seoul, Korea 136-701}
\address{$^{10}$INFN and University of Milano, Milano, Italy}
\address{$^{11}$University of North Carolina, Asheville, NC 28804}
\address{$^{12}$Dipartimento di Fisica Nucleare e Teorica and INFN, Pavia, 
Italy}
\address{$^{13}$University of Puerto Rico, Mayaguez, PR 00681}
\address{$^{14}$University of South Carolina, Columbia, SC 29208}
\address{$^{15}$University of Tennessee, Knoxville, TN 37996}
\address{$^{16}$Vanderbilt University, Nashville, TN 37235}
\address{$^{17}$University of Wisconsin, Madison, WI 53706}

\date{\today}
\maketitle
\begin{abstract}
Using a large sample of photoproduced charm mesons from the FOCUS
experiment at Fermilab (FNAL-E831), we observe the decay $D^0\!\rightarrow\! 
K^+\pi^-$ with a signal yield of $149\!\pm\!31$ events compared to a 
similarly cut sample consisting of $36\,760\!\pm\!195$ $D^0\!\rightarrow\! 
K^-\pi^+$ events.  We use the observed ratio of $D^0\!\rightarrow\!K^+\pi^-$ to 
$D^0\!\rightarrow\!K^-\pi^+$ $(0.404\pm 0.085\pm 0.025)\%$ to obtain a 
relationship between the $D^0$ mixing and doubly Cabibbo suppressed decay 
parameters.
\end{abstract}
 
\narrowtext

The decay $D^0\!\rightarrow\!K^+\pi^-$~\cite{cp} may occur either via a doubly 
Cabibbo suppressed (DCS) decay or through the mixing of the $D^0$ into 
$\overline{D}^0$ followed by the Cabibbo favored (CF) decay 
$\overline{D}^0\!\rightarrow\!K^+\pi^-$.  The na\"{\i}ve expectation for the 
ratio of DCS to CF branching fractions, $R_{\mathrm{DCS}}$, is 
$\tan^4\theta_C\!\simeq\!0.25\%$, which may be modified by final state 
interactions.  Contributions from nonperturbative long range interactions make 
exact calculations of Standard Model charm mixing difficult, but the rate is 
expected to be less than $10^{-3}$~\cite{Wolfenstein:1985ft,Donoghue:1986hh}.  
On the other hand, new physics may enhance
mixing~\cite{Nelson:1999fg}.  Since Standard Model charm 
sector $CP$ violation is expected to be small, and current searches report 
negative results~\cite{Link:2000}, we ignore $CP$ violation in this study.  
Conversely, recent charm sector mixing searches hint at an effect with a rate 
of order $10^{-4}$~\cite{Link:2000cu,Godang:1999yd}; hence possible mixing 
effects must be considered.

Four groups~\cite{Godang:1999yd,Cinabro:1993nh,Aitala:1998fg,Barate:1998uy} 
have studied the decay $D^0\!\rightarrow\! K^+\pi^-$ and measured its branching
ratio with respect to $D^0\!\rightarrow\! K^-\pi^+$, but only the most recent 
result~\cite{Godang:1999yd} by CLEO~II.V is statistically significant.  In 
addition, there is a variation of a factor of five among the measurements with 
the most recent yielding the most events and the lowest value.  We present a 
high statistics measurement of the branching ratio with more events than any 
previous experiment and different systematic uncertainties. 

The data were collected by the FOCUS Collaboration during the 1996-97 Fermilab 
fixed target run in the Wideband Photon beamline using an upgraded version of 
the E687 spectrometer~\cite{Frabetti:1992au}.  Charm particles are produced in 
the interaction of high energy photons ($\langle E \rangle\!\approx\! 180$~GeV) 
with a segmented BeO target.  In the target region, charged particles are 
tracked by 16 layers of silicon microstrip detectors which provide 
excellent vertex resolution.  The momentum of the charged particles is
determined by measuring their deflection in two oppositely polarized, large
aperture dipole magnets with five stations of multiwire proportional chambers.  
Particle identification is determined by three multicell threshold \v{C}erenkov 
detectors, electromagnetic calorimeters, and muon counters.

\section{Measurement Method}
To minimize systematic effects we apply the same selection algorithm to both 
$D^0\!\rightarrow\! K^+\pi^-$ and the normalizing mode, $D^0\!\rightarrow\! 
K^-\pi^+$.  We use the sign of the soft $\pi$ ($\tilde{\pi}$) from the decay 
$D^{*+}\!\rightarrow\!D^0\tilde{\pi}^+$ to tag the production flavor of the $D$ 
meson.  For the normalizing mode the charge of the $\tilde{\pi}$ is the same as 
the charge of the $D^0$ daughter $\pi$ (right sign, RS), but is opposite for 
the DCS or mixing mode (wrong sign, WS). 

A $D^0$ candidate consists of a pair of oppositely charged tracks that form a
decay vertex and have a $K\pi$ invariant mass between 1.7 and 
2.1~${\mathrm{GeV}}/c^2$.  The $D^0$ candidate is used as a seed to locate a 
production vertex consisting of at least two charged tracks in addition to the
$D^0$.  The production vertex is required to lie within $1\sigma$ of the 
nearest target material and be separated from the decay vertex by at least five 
times the separation error ($\ell/\sigma_{\ell}>5$).  Each vertex must have a 
confidence level (CL) greater than 1\% and the decay vertex tracks are required 
to be inconsistent with originating in the production vertex.   

Highly asymmetrical $K\pi$ pairs reconstructing with the $D^0$ mass are more
likely to be background than signal.  A cut is made on asymmetry, 
${\mathcal{A}}\!=\!|p_K\!-\!p_{\pi}|/|p_K\!+\!p_{\pi}|$, to reject these
candidates.  The $\mathcal{A}$ cut point is lowered linearly as the $D^0$
momentum decreases to achieve the best rejection of background.

For each charged track, the \v{C}erenkov particle identification algorithm 
generates a set of $\chi^2$-like variables, $W_i\!=\!-2\log$(likelihood), 
where $i$ ranges over the $e$, $\pi$, $K$, and $p$ hypotheses.  We define
$W_{\mathrm{min}}$ as the smallest $W$ of the four hypotheses and we
require $W_i\!-\!W_{\mathrm{min}}<4$, where $i$ refers to either the $K$ or 
$\pi$ hypothesis.  The $D^0$ daughters must also satisfy the slightly stronger 
$K\pi$ separation criteria of $\Delta W_{K}=W_{\pi}\!-\!W_K>0.5$ for the $K$ 
and $\Delta W_{\pi}=W_K\!-\!W_{\pi}>-2$ for the $\pi$.

Events with the decay $D^0\!\rightarrow\!K^-\pi^+$ where the $K$ has been
misidentified as a $\pi$ and the $\pi$ has been misidentified as a $K$, produce
false WS candidates.  These doubly misidentified events form a broad peak in 
the $K^+\pi^-$ mass distribution centered on the $D^0$ mass.  When a real 
$\tilde{\pi}$ tag is present, a peak indistinguishable from the real WS signal 
appears in the $D^*\!-\!D$ mass difference.  We treat this double 
misidentification background by imposing a hard \v{C}erenkov cut on the sum 
$\Delta W_K+\Delta W_{\pi}>8$, when the invariant mass of the $K\pi$ pair, with 
the $K$ and $\pi$ particle hypotheses swapped, is within $4\sigma$ of the $D^0$ 
mass.  From Monte Carlo (MC) studies of this cut, we expect double 
misidentification feed-through in the WS signal to be consistent with zero, and 
we determine a 90\% CL upper limit of 5\% of the observed WS yield.

All tracks assigned to the production vertex are considered as potential 
$\tilde{\pi}$ candidates.  The $\tilde{\pi}$ candidate must satisfy 
$W_{\pi}\!-\!W_{\mathrm{min}}\!<\!4$ and be inconsistent with being an electron
from a $\gamma$ conversion where the $\gamma$ comes from a $D^{0*}$ decay. 
This is achieved using information from the \v{C}erenkov system, 
electromagnetic calorimeters, and silicon microstrip detectors.

Significant fake contributions to the WS yield arise from the decay
$D^{*+}\!\rightarrow\!D^0\tilde{\pi}^+$ in which the $\tilde{\pi}$ was 
correctly reconstructed but the $D^0$ was not.  In the inset of 
Fig.~\ref{refl}a we show the combined contributions from MC generated decays 
$D^0\!\rightarrow\! K^+K^-$ and $D^0\!\rightarrow\!\pi^+\pi^-$, which are 
reconstructed as $K\pi$.  Both the $KK$ and $\pi\pi$ reflection peaks (below 
and above the $D^0$ mass respectively) have tails which extend into the $D^0$ 
mass region.  In the inset of Fig.~\ref{refl}b we plot the $K^+\pi^-$ mass 
contributions from MC events of all known $D^0$ modes, except two-body final 
states.  The flat background in the $D^0$ signal region is composed primarily 
of partially reconstructed and doubly misidentified $D^0$ decays to 
$K^-\pi^+\pi^0$ and $K^-\ell^+\nu$.  The mass difference plots 
(Fig.~\ref{refl}a and b) for reflected events in the $D^0$ signal region show 
peaked backgrounds in the $D^*$ signal region.  Using tighter particle 
identification to eliminate these backgrounds rejects about one third of our 
real signal events.  To avoid this we divide the RS and WS samples into 
1~${\mathrm{MeV}}/c^2$ wide bins in mass difference from 139 to 
179~${\mathrm{MeV}}/c^2$ and plot the $K\pi$ mass for each bin (a typical mass
plot is shown in Fig.~\ref{ex}).  The $K\pi$ mass distribution is then fit in 
each bin as follows: the structured reflections, $KK$ and $\pi\pi$, are fit 
using line shapes obtained with the MC, the unstructured background is fit by a 
degree two polynomial, and the $D^0$ signal is fit with a Gaussian.  By fitting 
in this way the real $D^0\!\rightarrow\!K\pi$ decays are isolated from the 
correlated $\tilde{\pi}$ backgrounds.  

The $D^0$ yields from the 80 $K\pi$ fits (40 each RS and WS) are plotted versus 
mass difference.  The two composite mass difference distributions, shown in 
Fig.~\ref{fit}, are fit for the WS to RS branching ratio ($R_{\mathrm{WS}}$).  
The background is fit using the following function:
\begin{equation}
f(\Delta m) = \alpha [(\Delta m-m_{\pi})^{1/2} + \beta (\Delta
m-m_{\pi})^{3/2}],
\end{equation}
where $\alpha$ and $\beta$ are fit parameters and separate 
parameters ($\alpha_{\mathrm{RS}}$, $\beta_{\mathrm{RS}}$, 
$\alpha_{\mathrm{WS}}$ and $\beta_{\mathrm{WS}}$ ) are used to fit the RS and 
WS distributions.  The shape of the high statistics RS signal is used to fit 
the WS signal.  In the WS $D^*$ signal region the full fit function is the sum 
of the WS background parameterization and the scaled RS signal, where the 
signal scale factor is $R_{\mathrm{WS}}$.  Modeling the WS signal in this way 
avoids complications arising from parameterizing the non-Gaussian signal.  We 
obtain $R_{\mathrm{WS}}\!=\!(0.404\pm0.085)\%$ and find $36\,760\!\pm\!195$ 
events above background in the RS signal region corresponding to a WS yield of 
$149\!\pm\!31$ events. 

Several sources of possible systematic errors were investigated, taking care
to avoid tests correlated to possible mixing effects.  Since the WS and RS 
modes are kinematically identical, we expect the systematic effects due to 
spectrometer acceptance and analysis cuts to cancel.

The effectiveness of the double misidentification cut was tested by watching 
the stability of $R_{\mathrm{WS}}$ as the \v{C}erenkov cut is varied and the
swapped-hypothesis mass window is widened.  We found no evidence of a 
systematic error due to doubly misidentified decays.

We investigated various methods of fitting the mass difference background 
such as using a different parameterization, and constraining the RS and WS 
shapes to be the same ($\beta_{\mathrm{RS}}\!=\!\beta_{\mathrm{WS}}$).  We 
looked for sensitivity to the MC reflection shapes by shifting the 
reflection distributions by $\pm2\ {\mathrm{MeV}}/c^2$, and we searched for a 
systematic dependence on the mass difference binning by shifting the bin 
centers and changing the bin widths.  All such tests returned values of 
$R_{\mathrm{WS}}$ consistent with the baseline measurement.  

To estimate the systematic error, measurements of $R_{\mathrm{WS}}$ were made 
with 136 different combinations of fit conditions and cut variations.  Each
measurement is assumed to be equally likely and we take the statistical 
variance of the measurements to be the systematic error on $R_{\mathrm{WS}}$.  
We obtain a systematic error of 0.025\%.

\section{$R_{\mathrm{DCS}}$ in the Presence of Mixing}
\label{mix}
The time dependent rate for WS decays relative to the CF branching fraction is 
\begin{equation}
\label{rate}
R(t) = \left[R_{\mathrm{DCS}} + \sqrt{R_{\mathrm{DCS}}}\ y^{\prime}t
+ \frac{(x^{\prime 2}\!+\!y^{\prime 2})}{4}t^2\right]e^{-t},
\end{equation}
where $t$ is in units of the $D^0$ lifetime.  We use the strong phase
($\delta$) rotated convention of CLEO~\cite{Godang:1999yd}, with
$y^{\prime}\!=\!y\cos{\delta}\!-\!x\sin{\delta}$ and 
$x^{\prime}\!=\!x\cos{\delta}\!+\!y\sin{\delta}$ where $x$ and $y$ are the 
usual mixing parameters, $x=\Delta m/\Gamma$ and $y=\Delta \Gamma/2\Gamma$.

Using a large MC sample (10 times the data) of 
$D^0\!\rightarrow\!K^-\pi^+$ decays generated with a pure exponential lifetime 
of 413~fs~\cite{Caso:1998tx}, we can calculate the expected number of WS events 
by reweighting each accepted MC event with a weight given by
\begin{equation}
\label{weight}
W_i\!=\!\frac{N_{\mathrm{data}}}{N_{\mathrm{MC}}}\!
\left(\!R_{\mathrm{DCS}}\! +\! \sqrt{R_{\mathrm{DCS}}}\ y^{\prime}t_i +\! 
\frac{x^{\prime 2}\!+\!y^{\prime 2}}{4}t_i^2\right),
\end{equation}
where $t_i$ is the generated proper time for event $i$, and $N_{\mathrm{data}}$ 
and $N_{\mathrm{MC}}$ are the number of accepted RS events in the data and 
MC\@.  Summing Eq.~\ref{weight} over all accepted MC events and dividing by 
$N_{\mathrm{data}}$ we obtain
\begin{equation}
\label{incl}
R_{\mathrm{WS}} = R_{\mathrm{DCS}} + 
\sqrt{R_{\mathrm{DCS}}}\ y^{\prime}\langle t\rangle +
\frac{x^{\prime 2}\!+\!y^{\prime 2}}{4} \langle t^2\rangle.
\end{equation}
The averages $\langle t\rangle$ and $\langle t^2\rangle$ are measured 
from the generated lifetime of the MC events accepted in the analysis.  
We find $\langle t\rangle\!=\!1.578\!\pm\!0.008$ and $\langle 
t^2\rangle\!=\!3.61\!\pm\!0.03$, where the errors are systematic, determined 
by comparing the reconstructed MC lifetime averages to the averages measured in 
data.  Using our measured value of $R_{\mathrm{WS}}$ we obtain an expression 
for $R_{\mathrm{DCS}}$ as a function of the mixing parameters $x^{\prime}$ and 
$y^{\prime}$.

In Fig.~\ref{true} we plot $R_{\mathrm{DCS}}$ as a function of $y^{\prime}$ 
for two values of $x^{\prime}$ that cover the CLEO~\cite{Godang:1999yd} 95\% CL 
of $|x^{\prime}|<0.028$.  For comparison, the mixing measurements of CLEO and 
FOCUS~\cite{Link:2000cu} are included in Fig.~\ref{true}.  The CLEO result 
comes from a direct measurement of $R_{\mathrm{DCS}}$, $x^{\prime}$ and 
$y^{\prime}$.  The FOCUS band represents a measurement of $y$ from lifetime 
differences between $CP$ even and $CP$ mixed final states.  The FOCUS $y$ 
measurement can only be directly compared to the other results in the case of 
$\delta\!=\!0$.  For a model dependent comparison of the CLEO and FOCUS direct 
mixing results see~\cite{bergmann:2000id}.

\section{Conclusion}
We observe a signal in the decay channel $D^0\!\rightarrow\! K^+\pi^-$ and 
measure its branching ratio relative to $D^0\!\rightarrow\! K^-\pi^+$ to be 
$(0.404\pm 0.085\pm 0.025)\%$.  If charm sector mixing is significant, 
the doubly Cabibbo suppressed component of the branching ratio can be 
determined from the measured ratio by using the relation expressed in 
Eq.~\ref{incl} and plotted in Fig.~\ref{true}.  If charm mixing is sufficiently 
small, the doubly Cabibbo suppressed branching ratio is simply equal to the 
measured ratio.  For comparison, Table~\ref{oldr} lists the existing 
measurements of this branching ratio, made with the assumption of no mixing.

We wish to acknowledge the assistance of the staffs of Fermi National
Accelerator Laboratory, the INFN of Italy, and the physics departments of the
collaborating institutions. This research was supported in part by the U.~S.
National Science Foundation, the U.~S. Department of Energy, the Italian
Istituto Nazionale di Fisica Nucleare and Ministero dell'Universit\`a e della
Ricerca Scientifica e Tecnologica, the Brazilian Conselho Nacional de
Desenvolvimento Cient\'{\i}fico e Tecnol\'ogico, CONACyT-M\'exico, the Korean
Ministry of Education, and the Korean Science and Engineering Foundation.

\bibliography{draft}

\begin{thebibliography}{10}

\bibitem{cp}
The charge conjugate mode is implicitly included.

\bibitem{Wolfenstein:1985ft}
L. Wolfenstein, Phys. Lett. {\bf B164},  170  (1985).

\bibitem{Donoghue:1986hh}
J.~F. Donoghue {\it et~al.}, Phys. Rev. {\bf D33},  179  (1986).

\bibitem{Nelson:1999fg}
For a compilation of Standard Model and Non-Standard Model theoretical 
predictions see H.~N. Nelson, e-Print Archive hep-ex/9908021, (1999). 

\bibitem{Link:2000}
J.~M. Link {\it et~al.}, Phys. Lett. {\bf B491},  232  (2000).

\bibitem{Link:2000cu}
J.~M. Link {\it et~al.}, Phys. Lett. {\bf B485},  62  (2000).

\bibitem{Godang:1999yd}
R. Godang {\it et~al.}, Phys. Rev. Lett. {\bf 84},  5038  (2000).

\bibitem{Cinabro:1993nh}
D. Cinabro {\it et~al.}, Phys. Rev. Lett. {\bf 72},  1406  (1994).

\bibitem{Aitala:1998fg}
E.~M. Aitala {\it et~al.}, Phys. Rev. {\bf D57},  13  (1998).

\bibitem{Barate:1998uy}
R. Barate {\it et~al.}, Phys. Lett. {\bf B436},  211  (1998).

\bibitem{Frabetti:1992au}
P.~L. Frabetti {\it et~al.}, Nucl. Instrum. Meth. {\bf A320},  519  (1992).

\bibitem{Caso:1998tx}
D.~E. Groom {\it et~al.}, Eur. Phys. J. {\bf C15},  1  (2000).

\bibitem{bergmann:2000id}
S. Bergmann {\it et~al.}, Phys. Lett. {\bf B486},  418  (2000).

\end{thebibliography}
\bibliographystyle{prsty}


\begin{figure}[b!]
\caption{Monte Carlo studies of  background contamination from (a) 
$D^0\!\rightarrow\!K^+K^-$ and $D^0\!\rightarrow\!\pi^+\pi^-$, and (b) all 
known multibody $D^0$ decay modes.}
\label{refl}
\end{figure}

\begin{figure}[t!]
\caption{A sample $K\pi$ mass fit.}
\label{ex}
\end{figure}

\begin{figure}[b!]
\caption{(a) The RS mass difference distribution, with the inset showing a 
close-up of the RS background fit and signal region.  (b) The WS mass 
difference distribution with the signal and background fit contributions 
shown.}
\label{fit}
\end{figure}

\begin{figure}[t!]
\caption{$R_{\mathrm{DCS}}$ plotted as a function of $y^{\prime}$.  Contours 
are given for two values of $x^{\prime}$ covering the 95\% CL of the CLEO II.V
result.} 
\label{true}
\end{figure}


\begin{table}
\caption{Measurements of $R_{\mathrm{DCS}}$ with the assumption of no charm 
mixing and no $CP$ violation.}
\label{oldr}
\begin{tabular}{lcc}
Experiment                     & $R_{\mathrm{DCS}}$ (\%) no Mixing &  Events\\ 
\hline
CLEO~\cite{Cinabro:1993nh}     & $0.77\pm0.25\pm0.25$               & 19.1  \\
E791~\cite{Aitala:1998fg}      & $0.68^{+0.34}_{-0.33}\pm0.07$      & 34    \\
Aleph~\cite{Barate:1998uy}     & $1.77^{+0.60}_{-0.56}\pm0.31$      & 21.3  \\
CLEO II.V~\cite{Godang:1999yd} & $0.332^{+0.063}_{-0.065}\pm 0.040$ & 44.8  \\ 
This Study                     & $0.404\pm 0.085\pm 0.025$          & 149 \\
\end{tabular}
\end{table}

\newpage
\centerline{
\includegraphics{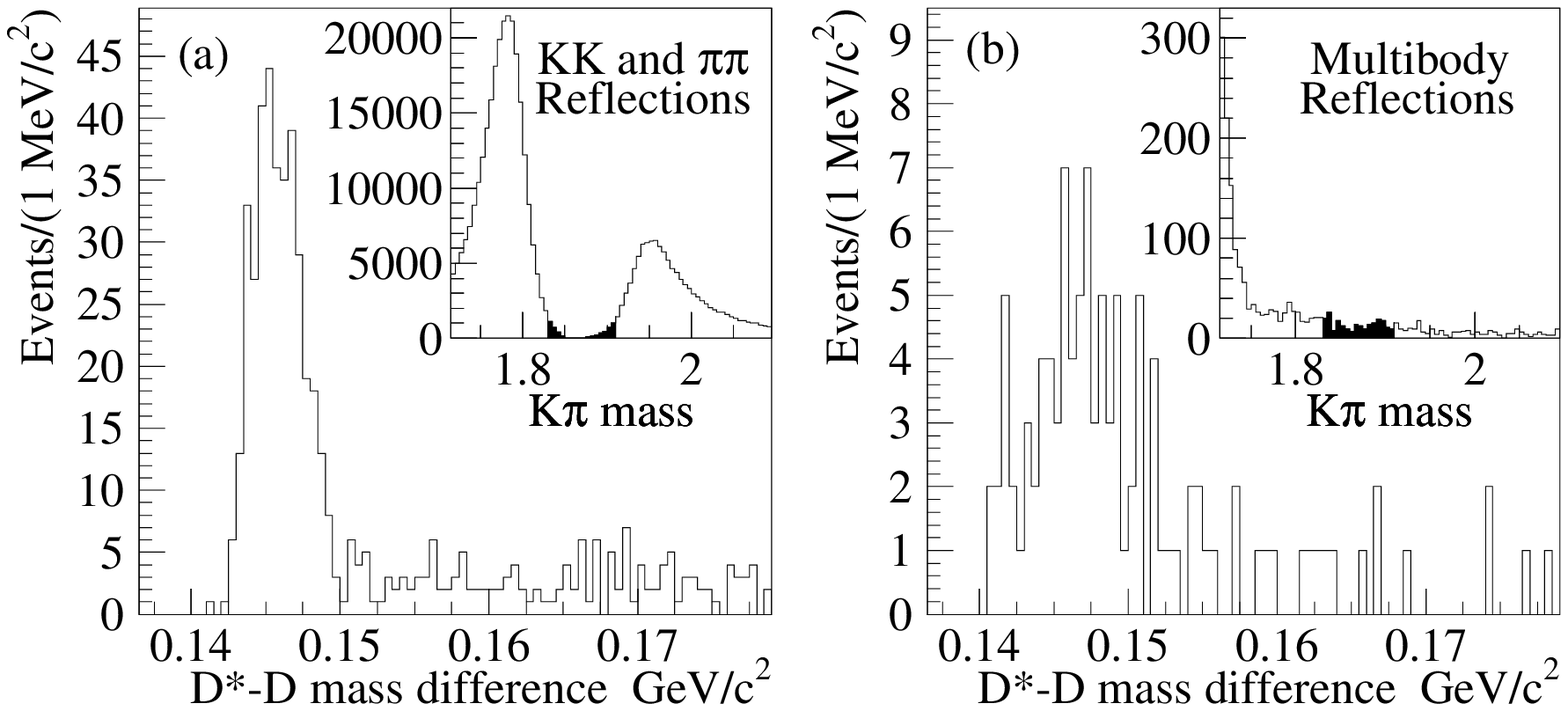}
}
\newpage
\centerline{
\includegraphics{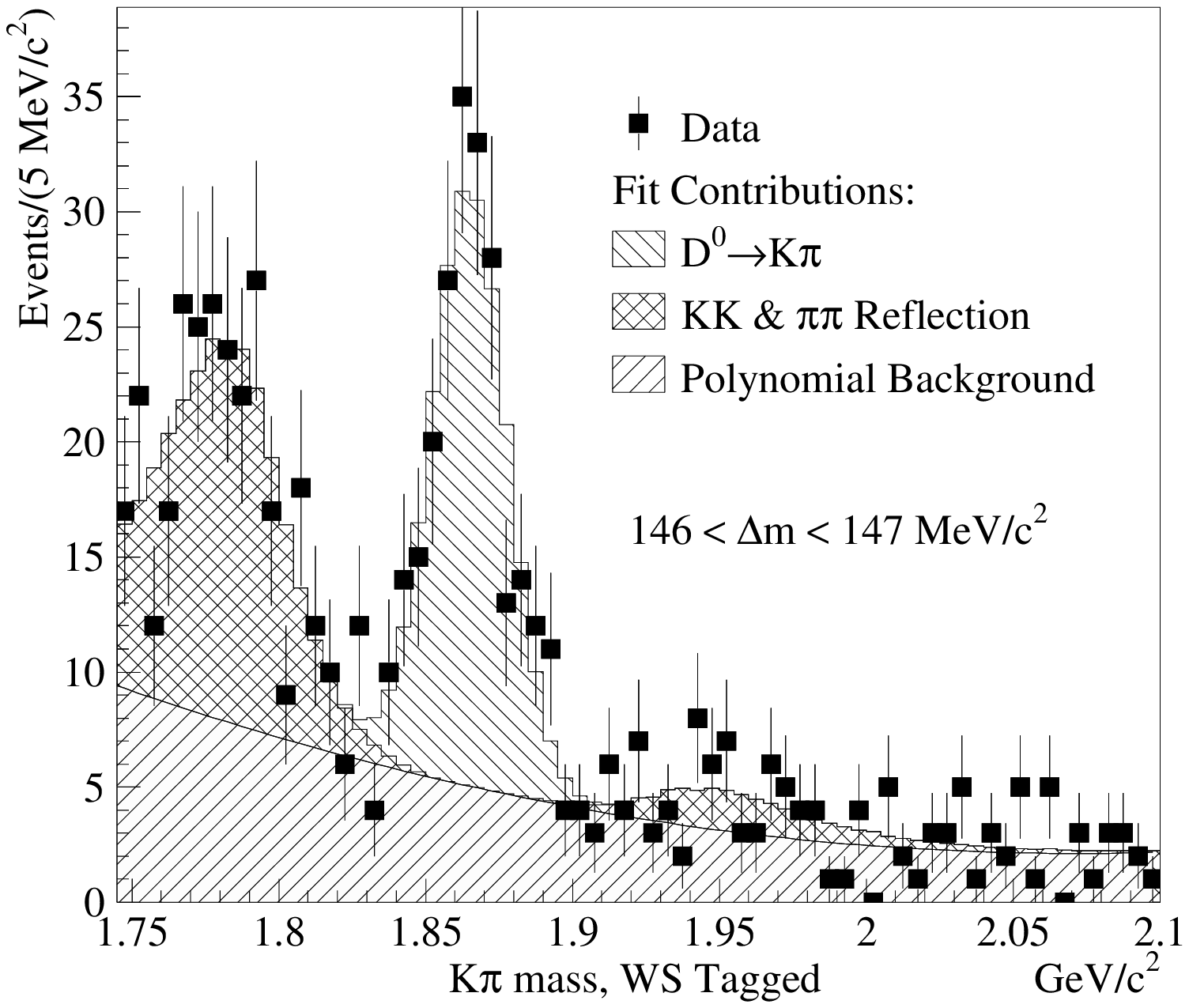}
}
\newpage
\centerline{
\includegraphics{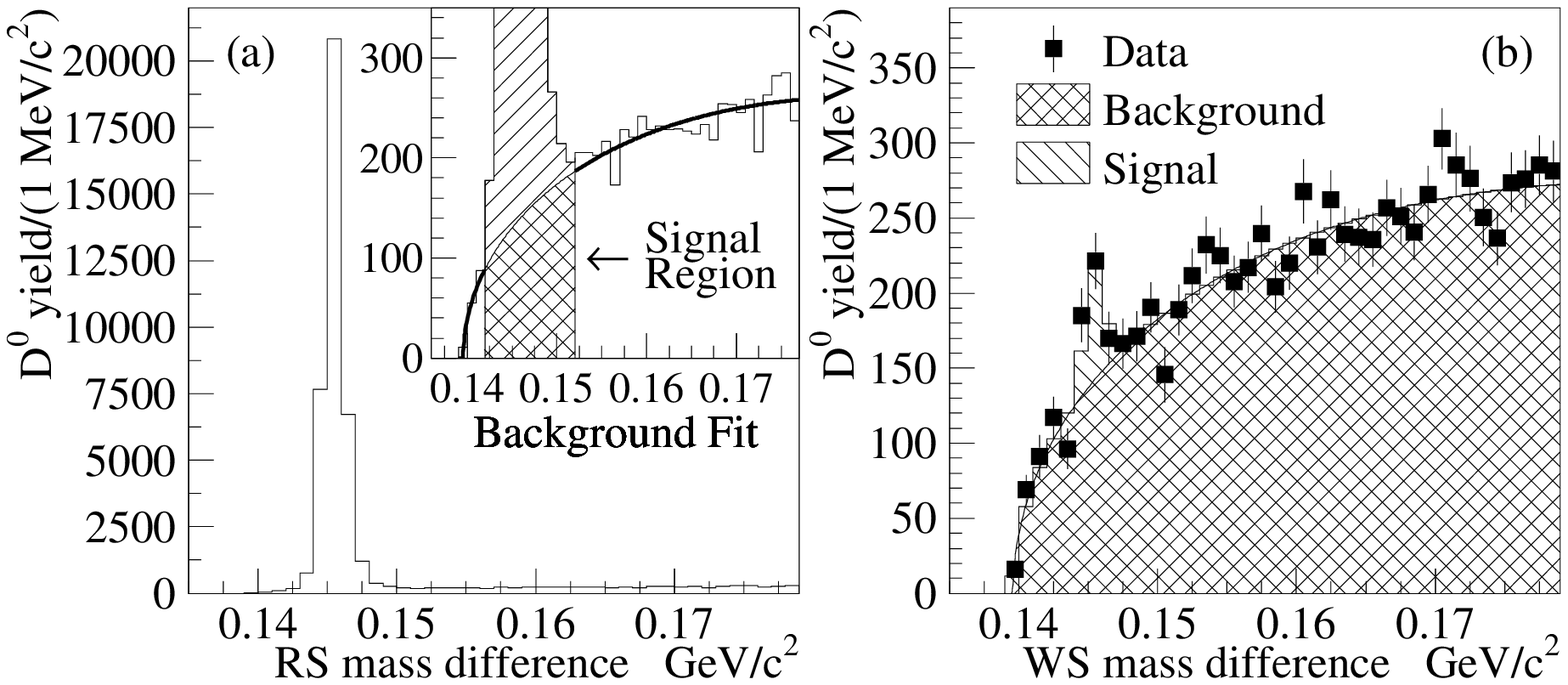}
}
\newpage
\centerline{
\includegraphics{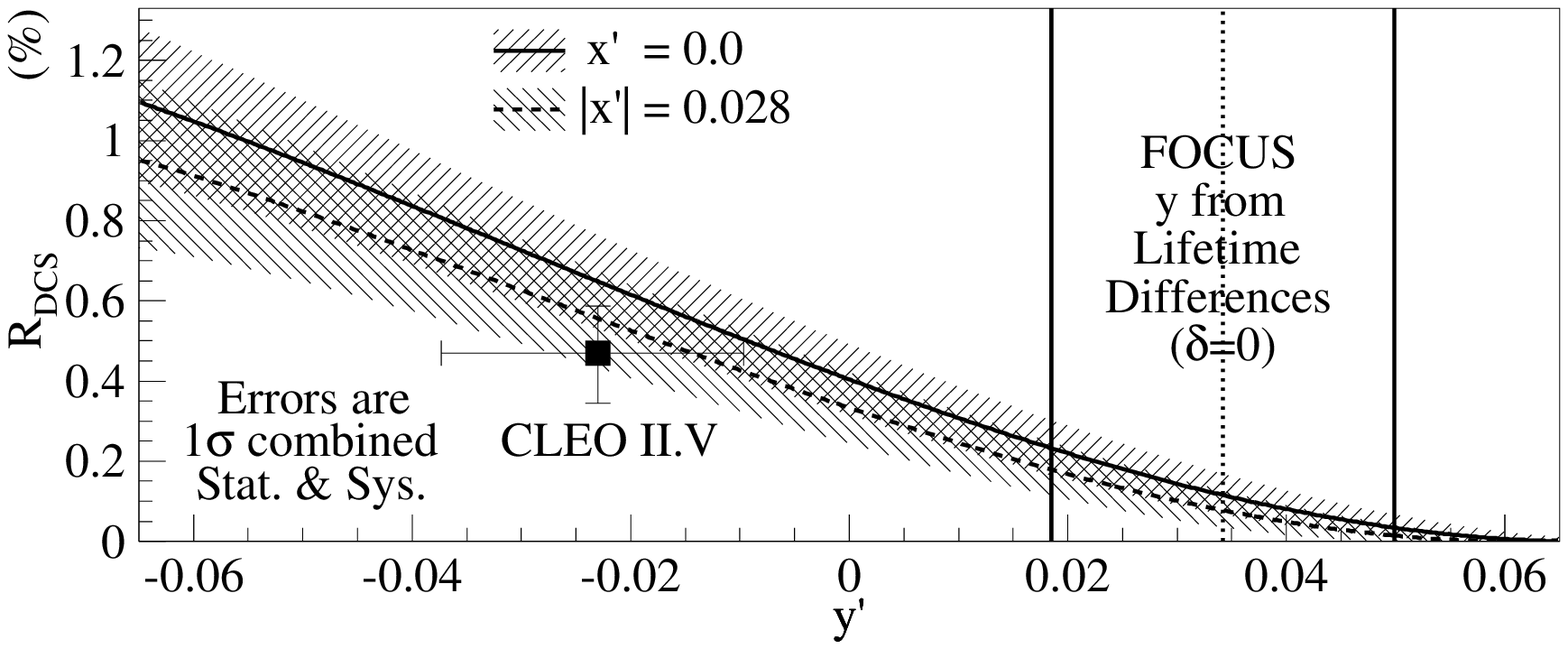}
}

\end{document}